     \documentstyle[12pt,world_sci]{article}
\begin{document}
    \def\DZ{D\O\ }
\def\METv{\mbox{${\Missing{\vec E}{T}}$}}
\def\ppbar{$p\bar p$ }
    \input macros
    \input psfig
    \input{EPSF}
    \input BoxedEPS.tex
    \SetRokickiEPSFSpecial
    \HideDisplacementBoxes
{\hfill AZPH-EXP/94-04}
{\hfill arch-ive/9412002}
     \title{W-Jet Rapidity Correlations in $p\bar p$ Collisions at
$\sqrt{s}=1.8$
     TeV\thanks{Invited Talk to the $24^{th}$
International Symposium on Multiparticle Dynamics, Vietri Sul Mare, Italy,
Septemeber 12 - 19, 1994} }
     \author{G.E. Forden\thanks{Representing the \DZ Collaboration}
     \\
     {\em Department of Physics\\ University of Arizona \\
     Tucson, AZ 85721}}
    \maketitle
     \setlength{\baselineskip}{2.6ex}

\begin{abstract} \DZ has used $W\rightarrow e\nu$ events produced
in association
with a high
$p_T$ jet to examine
the effects of strong radiative corrections.
We have compared the primary jet pseudorapidity distribution, as a
function of reconstructed W rapidity to leading order (LO) and next
to leading
order (NLO) QCD Monte Carlos, as well as a model based on extended
color dipoles.
Our preliminary analysis finds that the primary jet is more central
 than either
LO or NLO expectations and is in qualitative agreement with the
color
dipole model.
\end{abstract}

The transverse momentum of a charged vector boson, $W^\pm$,
produced in $p\bar p$ collisions arises from multiple gluon emission.
For sufficiently large transverse momentum, however, the dominant
production mechanisms are believed to be the annihilation and
Compton sub-processes
shown in Figure 1.
In both processes a single jet is
responsible for compensating the W's transverse momentum, and the
cross section
is maximal when the jet is at the same rapidity as the W.
This relationship will be altered by the proton's structure
function which determines the probability of finding suitable
initial partons to produce a given event topology.

This correlation between the jet's rapidity and the W's could be
affected by
several physical processes.  Gluons are
expected to be radiated preferentially between the jet and the
nearest beam.
The recoil from these jets
could systematically  shift the original jet to central
rapidities.  An
alternative process, based on the extended color dipole
model\cite{bo1},  would
preferentially produce
central gluon jets\cite{bo2} independent of the W's rapidity.

We present a study of the rapidity correlation of a high $p_T$
W boson and the primary jet as a function of W rapidity. We
used W events identified in the $e\nu$ channel by the \DZ  detector.
We do not distinguish  between the different W charges.
The resulting correlations were compared to the predictions from
lowest order\cite{LO} (LO) and
next to leading order\cite{NLO} (NLO) QCD
Monte Carlos, and
ARIADNE\cite{lonnblad},
a model based on
the extended color dipole.
The NLO Monte Carlo is an order $\alpha_s^2$ simulation.
This study is based
on the 14.9 $pb^{-1}$ taken during the 1992-93 Tevatron run.  The
\DZ  detector is described in detail elsewhere\cite{Abachi1}.  We
review here the features of the detector relevant for this
analysis.  The
$W\rightarrow e\nu$
events were selected by requiring at the hardware trigger level a
minimum of one electromagnetic (EM) trigger tower
$(\Delta\eta\times\Delta\phi
=0.2 \times 0.2)$ in the pseudo-rapidity range of $|\eta| \leq 3.2$
to have at
least 10 GeV transverse energy or two such towers with transverse
energy above
7 GeV.
The subsequent higher trigger levels require a cluster of EM cells
 with
a transverse energy of 20 GeV as well as some rudimentary shape
and isolation cuts.
Additionally, the level 2 software trigger
requires a missing $E_T$ for the event in excess of 20 GeV.  The
missing transverse energy vector,
\METv,
is calculated using the hit information from the entire calorimeter,
 EM as
well as hadronic, which covers pseudo-rapidity between $\pm 4$.
This trigger sequence has an over-all efficiency of 63\% ,
including the geometric acceptance.

\begin{figure}
\center{
\TrimTop{2.0in}
\TrimBottom{5.5in}
\TrimLeft{2.0in}
\TrimRight{1.0in}
\TrimBoundingBox{0.5in}
\hSlide{0.5in}
\ForceHeight{2.5in}\BoxedEPSF{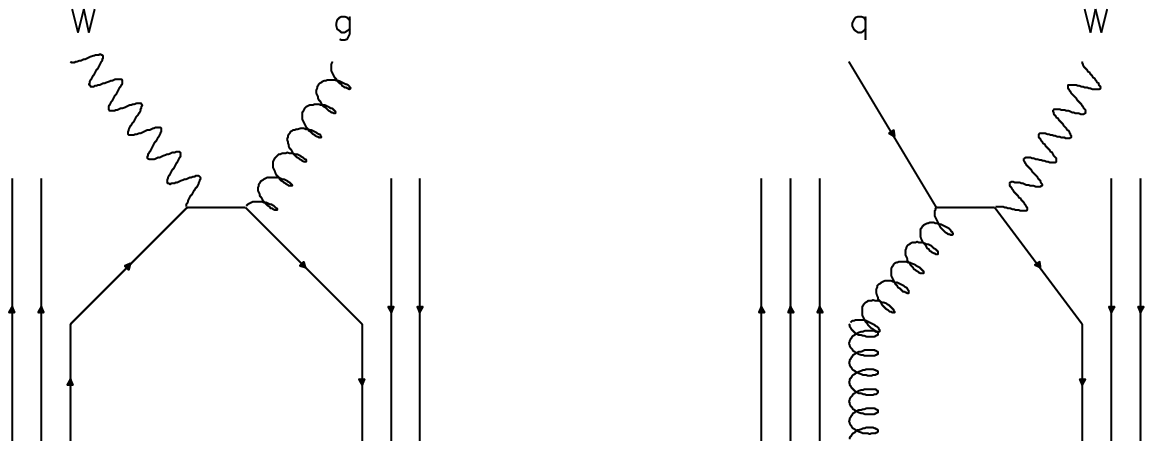}
}
  \caption{
    The two dominant lowest order processes, the annihilation
    and Compton diagrams, yielding high $p_T$ W bosons.
  }
\end{figure}

The off-line electron identification
requires that the candidate shower have 90\%  or greater EM energy
fraction
and that
an ``H-matrix" analysis\cite{hmatrix} of the
shower shape be consistent with an electron.
Furthermore, the candidate cluster must be isolated with $(E_{
.4}-EM_{.2})/EM_{.2} \le 0.15$.  The first term in the numerator,
$E_{.4}$,
is the  total energy (hadronic plus electromagnetic)
in a cone of $\Delta R= \sqrt{\Delta\eta^2+\Delta\phi^2}= 0.4$
centered on
the electron candidate.  The denominator, $EM_{.2}$, is the
electromagnetic energy inside the $\Delta R=0.2$ cone.
Finally, the
electron candidate is required to have a charged track pointing
toward the
shower centroid.
Only electrons
outside the region $1.1\le |\eta_{el}| \le 1.5$ and
with transverse energy greater than 25 GeV are considered when
forming W candidates.  The magnitude of \METv is also required
to be greater than 25 GeV.

The W boson candidates are then selected by requiring the
transverse mass formed by the electron candidate and the event's
\METv
to be greater than 45 GeV and less than 82 GeV.  This analysis
also requires
at least one jet with $p_T \ge 20$ GeV
and with pseudorapidity $(\eta)$ between -3 and
3.  The jets are found using a cone algorithm with $\Delta R=0.7$.
This jet also must
pass cuts designed to remove jet candidates resulting from spurious
hits
caused by showers associated with the main ring, as opposed to \ppbar
collisions in the Collider, etc.
These jet
quality cuts are 95\%  effective and remove only 4\%  of real jets.
The electron
isolation cut is further strengthened by requiring that the $\Delta R$
separation between any reconstructed jet and the electron candidate
is greater
than 1.3.

The transverse momentum of the W, $\vec P_\bot(W)$, is
determined by the event's missing transverse energy, \METv ,
plus the transverse momentum of the electron.  The W's momentum
component along the beam direction, $P_z$, is not directly measurable
since
a considerable fraction of the collision's energy escapes detection
down the
beam pipe.  However, $P_z(W)$    may be estimated for each event by
constraining the mass of the W candidate to be the world average mass,
80.22 GeV\cite{pdg}.
Knowing both the $P_z(W)$ and $\vec P_\bot(W)$
enables us to calculate the W's rapidity.
The W mass constraint gives, in general, two solutions for $P_z(W)$.
{\it A priori}, there are a number of ways of choosing the ``correct"
solution.  Monte Carlo studies indicate that
choosing the
minimum $|P_z|$ produces a good estimate.  This is the algorithm that we
use for this analysis.  Figure 2 shows the final sample
distributions of reconstructed
W rapidities from the final data sample as well as the LO plus background
(see below) and the ARIADNE plus background estimates.

\begin{figure}
\center{
\TrimTop{2.0in}
\TrimBottom{2.0in}
\TrimLeft{4.0in}
\TrimRight{1.0in}
\TrimBoundingBox{0.5in}
\hSlide{0.5in}
\ForceHeight{2.5in}\BoxedEPSF{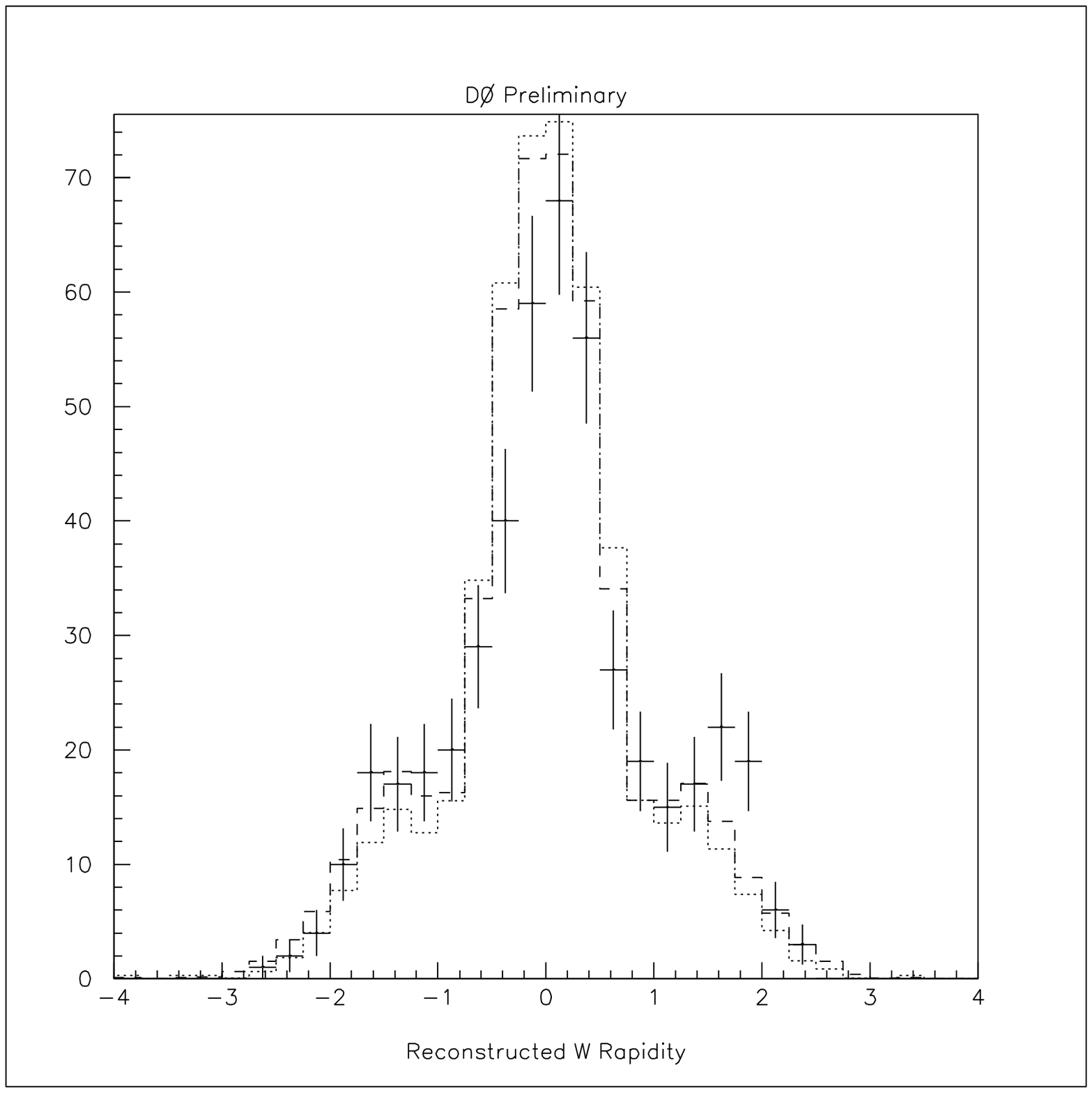}
}
  \caption{
The distribution of reconstructed W rapidities is shown together with the
color dipole (dashed line) and leading order (dotted) expectations.  The
Monte Carlos have had the background estimates included.
  }
\end{figure}

The background to $W\rightarrow e\nu+jets$ comes primarily from
a purely QCD
process where one of the jets fakes an electron.
We collected a sample of multijet events using our lowest jet $p_T$
trigger, which
at level 1
requires a single trigger tower with $E_T$ greater than 3 GeV.  These
events were then required to
satisfying all the kinematic
requirements for our W sample but not the electron quality cuts.
In selecting
this sample we
minimized the probability that a real electron would end up in this
background sample by requiring that the jet which will serve as the false
electron have a fraction of its total energy in the electromagnetic
calorimeter
between 0.1 and 0.7.  This background sample
will be referred to as the ``fake W" sample.  The fake W
sample is used to determine
the contributions from background events in the correlation distributions.
We divided the data and the fake W samples into six ranges of electron
$p_T$,
25 - 30, 30 - 35, 35 - 40, 40 - 45, 45 - 50, and 50 - 95 GeV.
We used the distributions of
the azimuthal angular differences between the primary jet and the
electron candidate, $\Delta\phi$, in each $p_T$ range to
determine the single ``scaling" factor for the background
and the single Monte Carlo scaling factor.  The six $\Delta\phi$
distributions
together with the resulting fits are shown in Figure 3.
These $\chi^2$ fits were performed over the entire $\Delta\phi$ range.
As our
final cut, we then imposed a requirement that $\Delta\phi \le 2.5$
for the
lowest $p_T$ electrons, those with $|p_T| \le 35$ GeV.  As an
example, the
electron-neutrino transverse mass distribution is shown in Figure 4
together
with the LO expectation for this distribution and the LO plus
background fit.

\begin{figure}
\center{
\TrimTop{2.0in}
\TrimBottom{1.5in}
\TrimLeft{4.25in}
\TrimRight{3.5in}
\TrimBoundingBox{0.5in}
\hSlide{0.5in}
\ForceHeight{6in}\BoxedEPSF{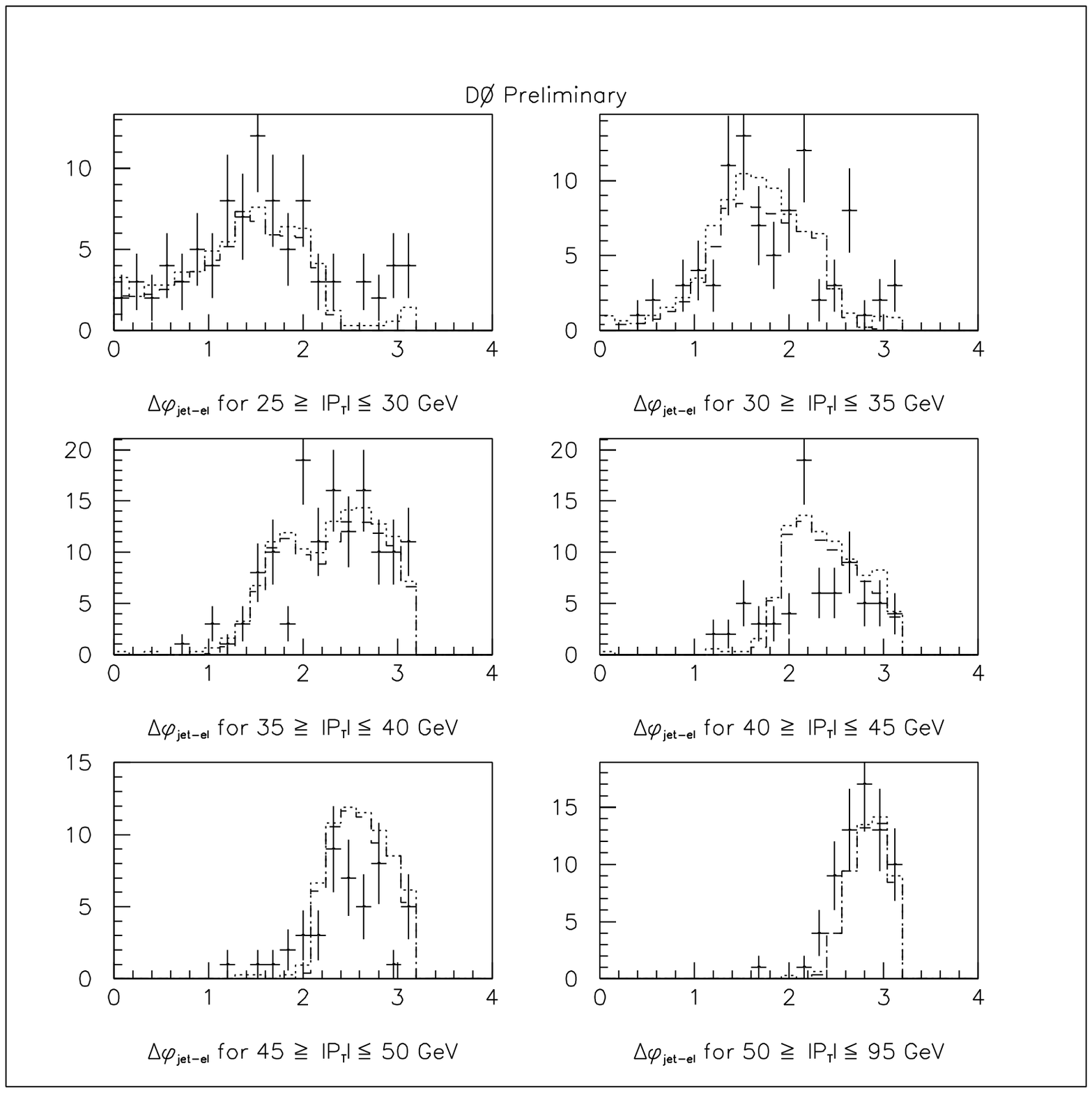}
}
  \caption{
The azimuthal angle differences between the electron candidate
and the primary jet
are shown for six intervals in electron transverse momentum.
These distributions
were used to determine the expected background fractions for
each Monte Carlo.  The leading
order predictions are shown here.
The dashed lines are the pure leading order and the dotted lines
are the LO plus background.
  }
\end{figure}

\begin{figure}
\center{
\TrimTop{2.0in}
\TrimBottom{2.0in}
\TrimLeft{4.0in}
\TrimRight{1.0in}
\TrimBoundingBox{0.5in}
\hSlide{0.5in}
\ForceHeight{2.5in}\BoxedEPSF{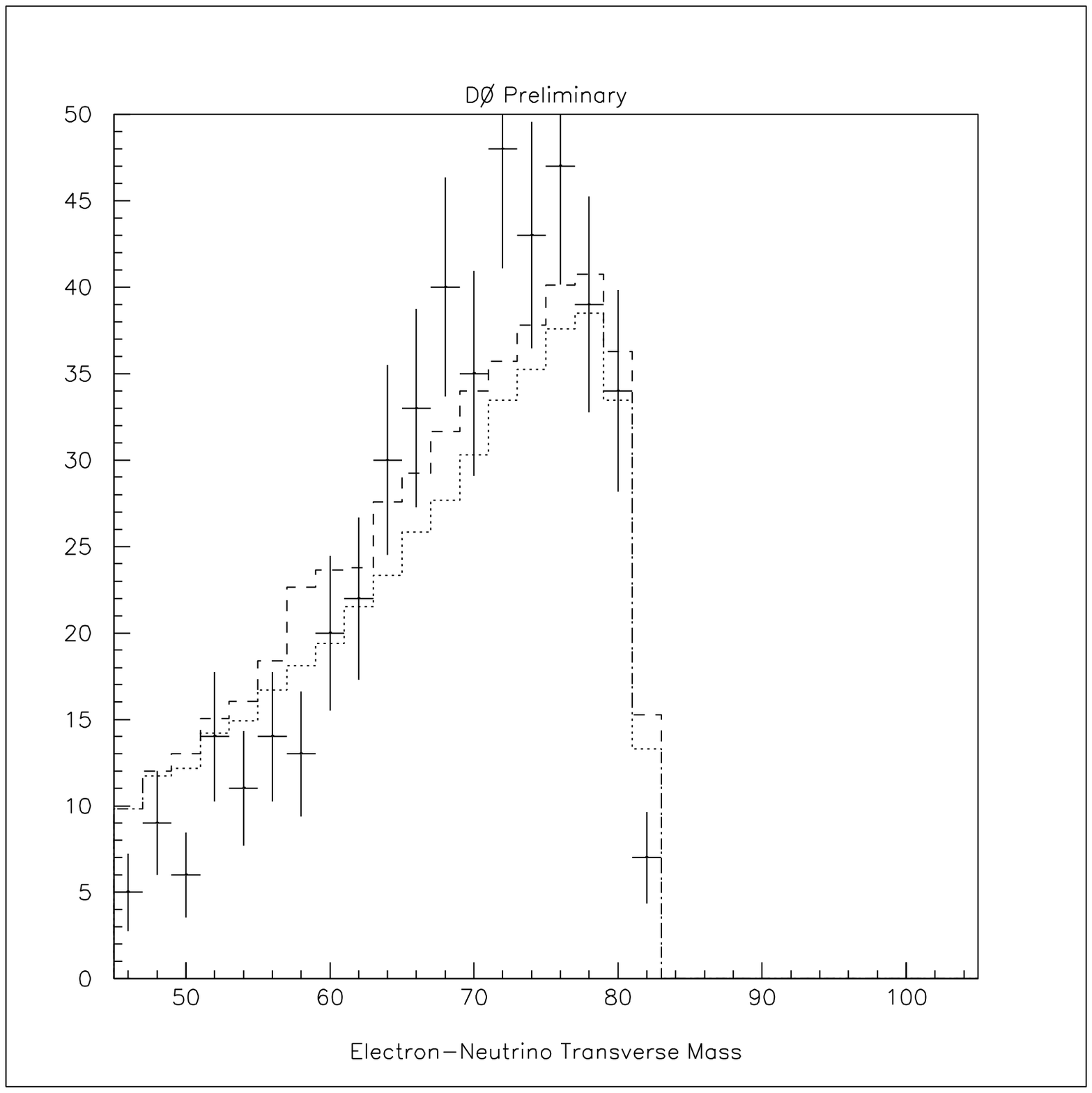}
}
  \caption{
The electron neutrino transverse mass distribution.  The leading
order prediction
(dotted line) and the
leading order plus background (dashed line) are shown for
comparison.  The
background fraction was determined using the $\Delta \phi$
distributions as
shown in Figure 3.
  }
\end{figure}

The final data sample of 470 events (with an estimated background
of 45 events)
was divided into five subgroups according to
the reconstructed W rapidity, $|y_w|$, between 0 and 2.5.  In each
event, the
sign of each jet's pseudorapidity was determined by $\eta_{jet}
y_w/|y_w|$.  This has the effect of assigning positive pseudorapidity
to jets on
the same side as the reconstructed W and negative pseudorapidity to
those jets
opposite, in $\eta$, of the reconstructed W rapidity.
The primary jet is chosen by considering all those jets with
$\vec P_T^{jet} \cdot \vec P_T^W \le 0$ and then picking the one whose
$|\vec P_T^{jet}|$
is closest to the W's transverse momentum.
Similar procedures were done for each of the models considered except
that the corresponding backgrounds to each Monte Carlo were added and
then the
average was determined.  The resulting distributions are shown in
Figure 5.  The
data remains more central than the LO plus background expectations.
The NLO
plus associated background is more central than the LO predictions,
as was to be
expected from angle ordering considerations, but is still not as
central as the
data.  The color dipole model, ARIADNE, is in some what better agreement
with the data.

\begin{figure}
\center{
\TrimTop{2.0in}
\TrimBottom{1.5in}
\TrimLeft{4.25in}
\TrimRight{3.5in}
\TrimBoundingBox{0.5in}
\hSlide{0.5in}
\ForceHeight{6in}\BoxedEPSF{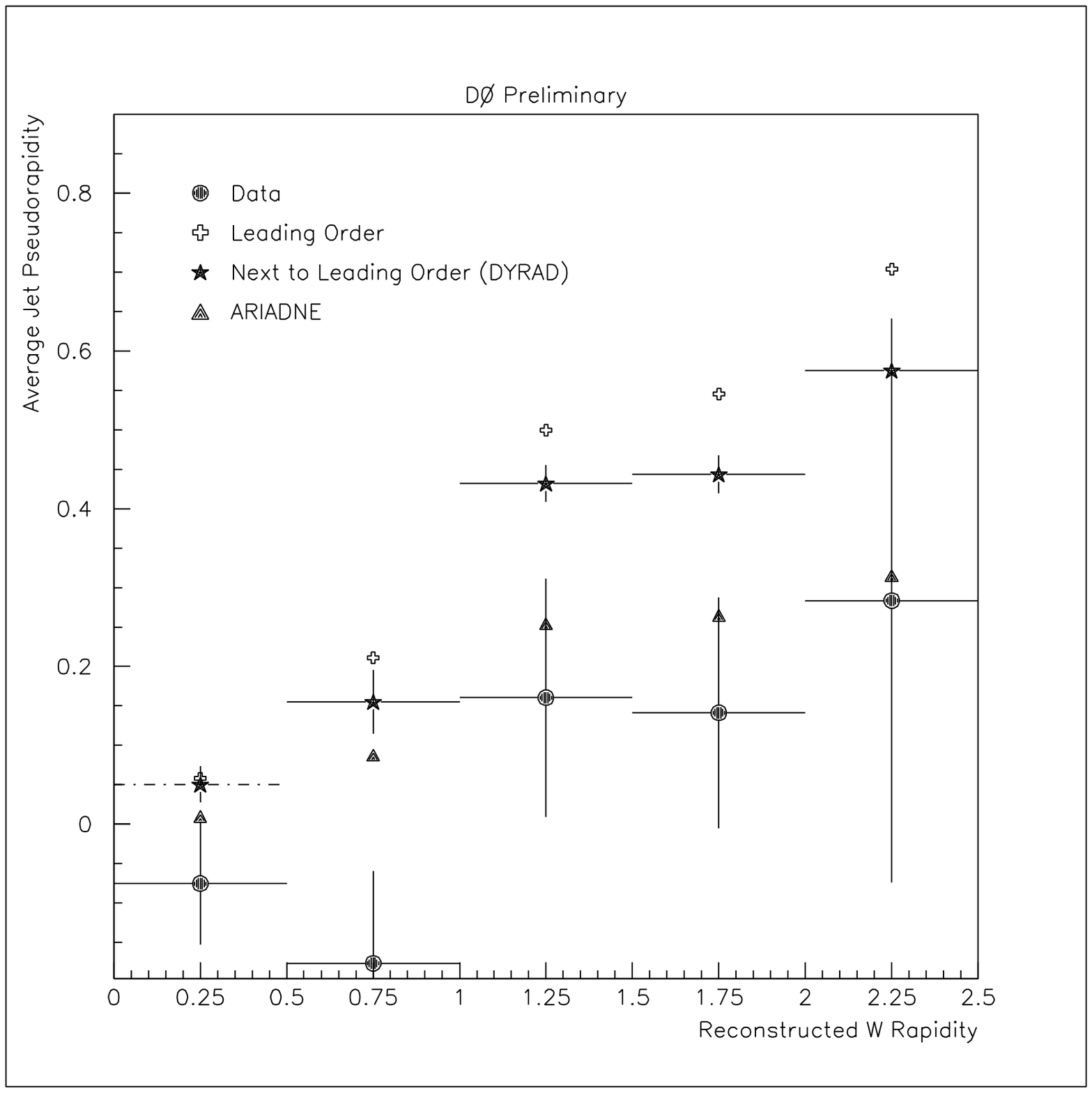}
}
  \caption{
The average jet pseudorapidity for the data and the various Monte
Carlos are shown.
The Monte Carlo's have had the background estimates included in
their averages.
Positive jet pseudorapidity is defined as $\eta_{jet}^{lab} y_w/|y_w|$,
so positive
rapidity in this plot implies that the jet is on the same side as the W.
  }
\end{figure}

We varied the structure functions
for the LO and NLO estimates using various modern sets;
Morfin-Tung LO\cite{MT}
and
CTEQ2L\cite{cteq}
for the LO and
CTEQ2M and
KMRSB0\cite{kmrs}
for the NLO.
We saw only insignificant differences.
We have also varied the energy scale for the LO structure functions
between $M_T^2$ and $M_T^2/4$ without detecting a noticeable effect.
This
insensitivity to the proton structure is expected given the large
$Q^2 \approx M_W^2$ for these events.
We ruled out $\eta$ symmetric jet
reconstruction inefficiencies by
considering the ratio
of jets on the same side as the W versus jets on the opposite side.
Asymmetric
$\eta$
inefficiencies were ruled out by considering the production of jets
in positive
lab frame $\eta$ (here defined as the direction of the incident proton)
to those with negative lab frame $\eta$.
These studies have rule out any significant $\eta$ dependent
reconstruction
inefficiencies.
We have also tried
changing our definition of ``primary" jet to be the jet with highest
$p_T$ with
no noticeable effect.  It should be pointed out that the W's $p_T$
is highly
correlated with the primary jet from
either of these definitions.  This implies that additional radiation
is
not playing a major role.

Finally, we have examined the dependence of our measurements on the
algorithm for
determining the rapidity of the W.
We have tried both unfolding these mistakes on
the data directly and using a weighting scheme\cite{marty} without
noticeable
changes in our final distributions.  Figure 6 shows the distributions
of
average jet rapidities for the NLO Monte Carlo using the reconstructed W
rapidity and the generated W rapidity.
In this preliminary analysis there is a systematic shift towards lower jet
rapidities.
However, these shifts
do not account for the
differences between the data and the various Monte
Carlo expectations.

\begin{figure}
\center{
\TrimTop{2.0in}
\TrimBottom{1.5in}
\TrimLeft{4.0in}
\TrimRight{1.0in}
\TrimBoundingBox{0.5in}
\hSlide{0.5in}
\ForceHeight{3.in}\BoxedEPSF{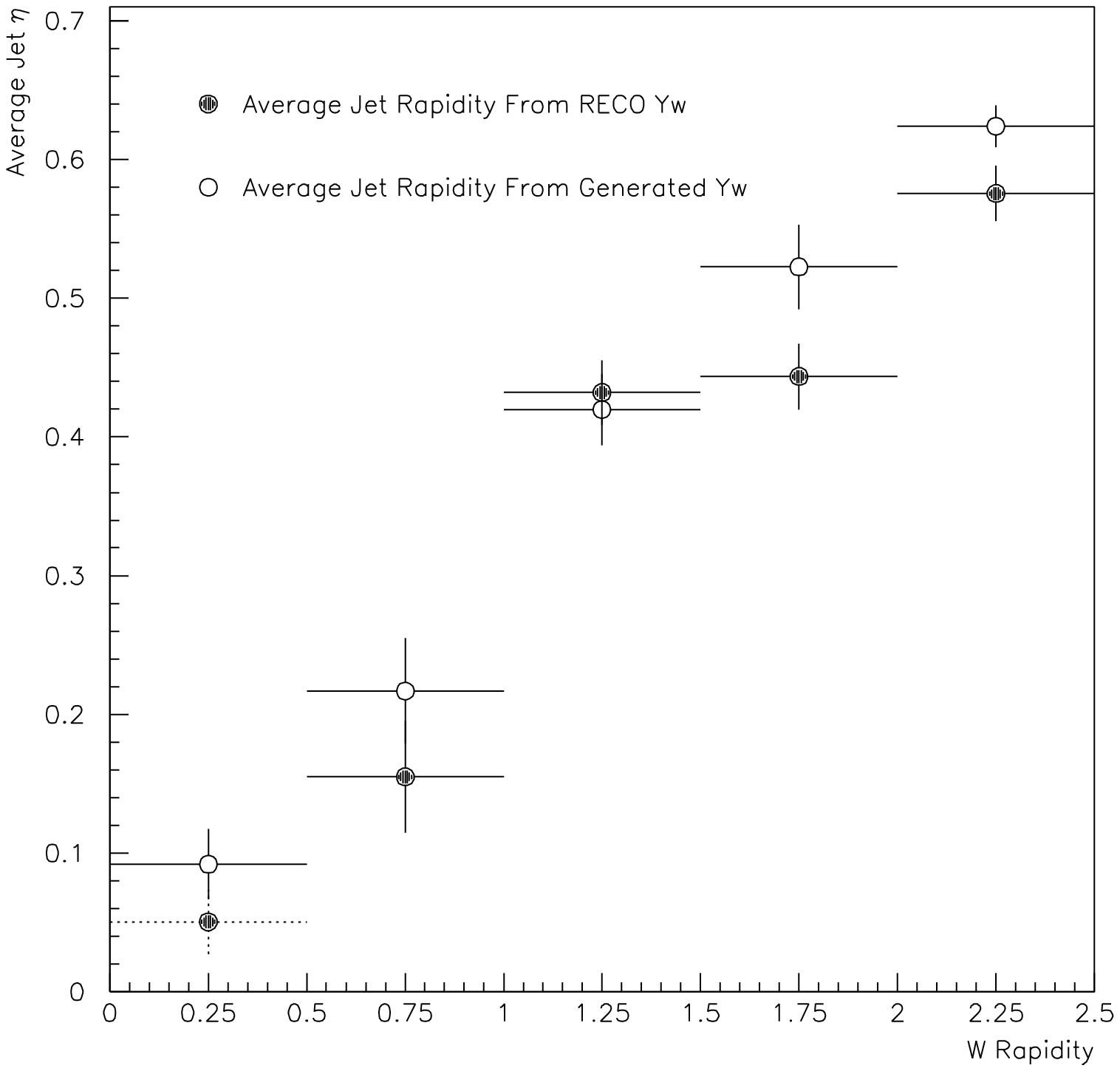}
}
  \caption{
The average jet pseudorapidity for the NLO Monte Carlo is shown as a
function of
the W's rapidity calculated using either the reconstructed W rapidity
 (solid dots)
or the rapidity at which the W was generated (open circles).
  }
\end{figure}

We have studied the rapidity correlation between the reconstructed W
and the
``primary" jet.
None of the models for high $p_T$ W production in \ppbar  collisions
do a
particularly good job of reproducing the observed behavior of the data
 in that
the primary jet remains central, independent of the reconstructed W's
rapidity.  The model based on extended color dipoles does the
best job of predicting the W-jet correlation.
The extended color dipole model assumes that all the color charges
inside the incident proton
contribute to determining the radiation pattern for the primary jet.

We would like to thank B. Andersson whose many interesting
conversations stimulated this study.  We are also grateful to L.
L\"onnblad for
providing us the ARIADNE Monte Carlo program for simulating high
$p_T$ W
events.  We particularly wish to express our gratitude to the
technical staff
at
Fermilab for their excellent support.  Financial support has
been provided
by the U.S. Department of Energy, the State Committee for Atomic
Energy
in Russia, the Comissariat a L'Energie Atomique in France, the U.S.
National Science Foundation, the Department of Atomic Energy in India,
CNPq in Brazil, the CONACyT in Mexico, and Colciencias in Colombia.

\pagebreak[4]


\begin{thebibliography}{99}
\bibitem{bo1} B.Andersson, G. Gustafson, L. L\"onnblad, U. Pettersson,
Z. Phys. C {\bf 43}, 625-632 (1989)
\bibitem{bo2} B. Andersson, Private Communication
\bibitem{LO} I. Hinchliffe LBL-34372, Submitted to Workshop on Physics
at Current Accelerators and the Supercollider, Argonne, IL, 2-5 Jun 1993
\bibitem{NLO} W.T. Giele, E.W.N. Glover, D. A. Kosower Phys. Lett.
B{\bf 309}
205 (1993)
\bibitem{lonnblad} L. L\"onnblad, Comp. Phys. Comm. {\bf 71} 15 (1992).
We
have used ARIADNE version 4.05P02 in this analysis.
\bibitem{Abachi1} S. Abachi {\it et al.,} Nucl. Instr. and Meth. in
Phys. A{\bf
338}
185 (1994)
\bibitem{hmatrix} R. Engelmann Nucl. Instr. Meth. {\bf 216} 45 (1983);
M. Narain, $7^{th}$ Meeting of the APS Divisions of Particles and
Fields, C.
H. Albright ed.  World Scientific, Batavia Il. (1992)
\bibitem{pdg} K. Hikasa {\it et al.,} Phys. Rev. D {\bf 45} (1992)
\bibitem{MT} J. Morfin and W.K. Tung, Z. Phys. C{\bf52} 13 (1991)
\bibitem{cteq} J. Botts {\it et al.}, Phys. Lett. {\bf 304}B 159 (1993)
\bibitem{kmrs} J. Kwiecinski, A.D. Martin, R.G. Roberts and W.J. Stirling,
Phys. Rev. D{\bf 42} 3645 (1990)
\bibitem{marty}  We would like to thank M. Block for making this
suggestion
during the workshop.
\end{thebibliography}
\end{document}